\documentclass[onecolumn,floatfix,superscriptaddress,showpacs,nofootinbib,preprint]{revtex4}
\usepackage{epsfig}
\usepackage{amssymb,latexsym,amsmath}

\newcommand{\be}{\begin{equation}}
\newcommand{\ee}{\end{equation}}
\newcommand{\bea}{\begin{eqnarray}}
\newcommand{\eea}{\end{eqnarray}}

\begin{document}

\title{Direct photon production from hadronic sources
in high-energy heavy-ion collisions}

\author{E.L.~Bratkovskaya}
\affiliation{Frankfurt Institute for Advanced Studies, Frankfurt, Germany}
\author{S.M. Kiselev}
\affiliation{Institute for Theoretical and Experimental Physics,
Moscow, Russia}
\author{G.B. Sharkov}
\affiliation{Institute for Theoretical and Experimental Physics,
Moscow, Russia}

\begin{abstract}
The low $p_T$ direct photon production from a variety of the hadronic
sources is studied within the microscopic HSD transport approach for
$p+C, p+Pb$ and $Pb+Pb$  collisions at 160 A GeV.  The
direct photon emission from elementary hadronic scatterings as well
as meson-meson bremsstrahlung are incorporated. The influence of
in-medium effects such as a collisional broadening of the vector-meson
spectral functions on the photon emission rate is found to be hardly
observable in the final spectra which are dominated by
bremsstrahlung type processes.  The uncertainties in the subtraction of
the 'background' from the photon decay of hadronic resonances inside
the hot and dense fireball is investigated, additionally. Our findings
are relevant for the interpretation and extraction
of experimental data on direct photon production at low $p_T$.
\end{abstract}


\maketitle


\section{Introduction}

The properties of hadronic matter under extreme densities and temperatures
and their phase transition to the deconfined and strongly interacting
Quark-Gluon-Plasma (QGP)  are the central topics of modern high-energy
physics. In order to understand the dynamics and relevant
scales of this transition laboratory experiments under controlled
conditions are presently performed with ultra-relativistic
nucleus-nucleus collisions. The electromagnetic radiation
(real and virtual photons, i.e. dilepton pairs) is the unique
probe to study the heavy-ion collisions since the photons
do not suffer from the final state interactions with the
surrounding matter and, thus, provide a clear signal of the various
environments of their creations
(cf. the pioneering works \cite{Fei76,Shu78,Bjo76}).
Moreover, the photons are emitted from all stages of the collisions --
from the  QGP  to the hadronic phase, so they may provide information
about the different degrees of freedom - from partons to hadrons
\cite{Pis89,Bra90,Kap91,epj:sri99}.

On the quark-gluon level the main processes for the direct photon
production are Compton scattering of quarks and gluons $q g \rightarrow
q \gamma$ and annihilation of quark and anti-quarks $q \bar q
\rightarrow g \gamma$ as leading processes, while the next leading order
(NLO) process is dominated by bremsstrahlung $q g \rightarrow q  g
\gamma$.  The photons radiated from the different sources have a
different transverse momentum $p_T$ that allows to separate the
contributions:  The photons with high $p_T$ are primarily produced by
the initial hard NN collisions and called as 'prompt' (or 'hard')
photons. Direct photons from a thermalised QGP dominate at lower
transwerse momentum ($1\le p_T\le 3$ GeV/c) and denoted as 'thermal
photons'. On the hadronic level the main source of direct photons is
meson-meson rescattering:  $\pi \rho \rightarrow \pi \gamma$, $\pi \pi
\rightarrow \rho \gamma$, $\pi K \rightarrow K^{*} \gamma$, $K \rho
\rightarrow K \gamma$, ...,  where the first two channels
are the most important. These photons from the hadronic rescattering
dominate at even lower $p_T$.
Additionally there are photons coming from the hadron decays (such as
$\pi^0, \eta$ etc.), which give a huge photonic 'background' and make
the experimental measurement of direct photons very complicated.

Most of the theoretical predictions for direct photons are based on the
local thermalization assumption and evaluate the photon production
rates from the equilibrated quark-gluon plasma or hadronic matter which
then are convoluted with the space-time evolution of the system.
In order to account for the non-equilibrium dynamics with its full complexity
one needs microscopic transport models. However, most of the
transport models -- commonly used for the description of high energy
heavy-ion collisions -- are based on the hadron-string picture
(e.g. HSD\footnote{ Hadron-String-Dynamics transport approach}
and UrQMD\footnote{Ultra-relativistic-Quantum-Molecular-Dynamics})
and do not include the phase transition from partonic to hadronic
matter in a consistent way, nor solve the hadronization problem. We
mention that there are attempts to develop such transport models,
e.g. the multi-phase transport model (AMPT) \cite{Ko_AMPT} which
includes pQCD-like partonic scattering, or combined models -
hydrodynamics for the QGP + UrQMD for the hadronic stage
\cite{UrQMD_hydro}, or Parton-Hadron-String-Dynamics (PHSD) model which
is based on a dynamical quasiparticle model  matched to reproduce
lattice QCD results in thermodynamic equilibrium \cite{PHSD_DQPM}.

We recall that already in 1995-1997 there were attempts
do describe the first experimental photon data by WA98 Collaboration
on S+Au at 200 A GeV \cite{WA98_SAu} in the transport models -
UrQMD \cite{Dumitru95}, AMPT \cite{AMPT_gam} and HSD \cite{Brat97}.
The transport calculations included the photon production by the hadronic
decays \cite{Dumitru95,AMPT_gam,Brat97} as well as direct photon
production from the meson-meson scattering \cite{Dumitru95,AMPT_gam}.
Also the influence of in-medium effects such as dropping vector meson
masses on the photon yield have been addressed in  Refs.
\cite{AMPT_gam,Brat97}. Unfortunately, the first WA98 data
provided only an upper bound for the direct photon production yield
and did not allow to draw solid conclusions  -- all transport results
were just below the upper limit given by WA98.
In 2000 the WA98 Collaboration provided  new data for  Pb+Pb at
160 A GeV \cite{WA98data}. This stimulated a new wave of interest
for direct photons from the theoretical side (cf. the
review~\cite{PhysRep2002}) and references therein).

The aim of our present study is to investigate the photon production at
SPS energies from hadronic sources using the extended version of the
HSD transport model \cite{Brat07off,Brat08SPS} which includes the
off-shell dynamics of the vector mesons with dynamical  spectral
functions. It allows  to study the influence of in-medium effects
-- such as a collisional broadening -- on the photon emission  rate.
We will compare our transport calculations to the experimental
data for Pb+Pb at 160 A GeV as well as to the new preliminary data on
photon production in p+C and p+Pb collisions at 160 A GeV
\cite{WA98_pA08}.  We stress here again that the HSD model doesn't
include the phase transition from QGP to hadronic matter;  we thus
concentrate on the low $p_T$ photon production which is dominated by
hadronic sourses.  First 'pilot' HSD results have been reported in Ref.
\cite{Kiselev}.

Our paper is organized as follows:
In Section II we describe the treatment of various channels for the
photon production in HSD. Section III contains a comparison
of the HSD results with the WA98 data for Pb+Pb at 160 A GeV, while
Section IV contains our results for  photon production in p+C and
p+Pb collisions at 160 A GeV. A summary closes this work in Section V.

\section{Photon production in HSD}

Our analysis is carried out within the HSD transport model
\cite{Brat97,CBRep98,Ehehalt} - based on covariant self energies
for the baryons \cite{KWeber} - that has been used for the
description of $pA$ and $AA$ collisions from SIS to RHIC energies.
We recall that in the HSD approach nucleons, $\Delta$'s,
N$^*$(1440), N$^*$(1535), $\Lambda$, $\Sigma$ and $\Sigma^*$
hyperons, $\Xi$'s, $\Xi^*$'s and $\Omega$'s as well as their
antiparticles are included on the baryonic side whereas the $0^-$
and $1^-$ octet states are incorporated in the mesonic sector.
Inelastic baryon--baryon (and meson-baryon) collisions with energies
above $\sqrt s_{th}\simeq 2.6$~GeV (and  $\sqrt{s_{th}} \simeq 2.3$~GeV) are
described by the FRITIOF string model \cite{FRITIOF} whereas low
energy hadron--hadron collisions are modeled in line with experimental
cross sections.
Note that the HSD transport approach includes the off-shell dynamics
of vector mesons explicitly;  for the details we address the reader to
Ref.  \cite{Brat07off}.

We consider the following hadronic sources of photon production:

I. The photon production by a mesonic decays
($\pi^0, \eta, \eta^\prime, \omega, \phi, a_1$) where
the mesons are produced firstly in baryon-baryon ($BB$), meson-baryon
($mB$) or meson-meson ($mm$) collisions.
The photon production from the mesonic decays represents a
'background' for the search of the direct photons, however, this
background is very large relative to the expected direct photon signal.
Moreover, there are a severe experimental difficulties in subtracting
the photons from hadronic decays.  For the present study we consider
the contributions from the photon decay of the following mesons:
\begin{eqnarray}
&&\pi^0 \to \gamma+ \gamma, 		\label{decR} \\
&& \eta \to \gamma + \gamma,         \nonumber\\
&&\eta^\prime \to \rho + \gamma,    \nonumber\\
&&\omega \to \pi^0 + \gamma,        \nonumber\\
&& \phi \to \eta + \gamma,           \nonumber\\
&& a_1 \to \pi + \gamma.             \nonumber
\label{l}\end{eqnarray}
The decay probability is calculated according to the corresponding
branching ratios taken from PDG \cite{PDG}. The broad resonances in the
initial/final state are treated in line with to their in-medium
spectral functions which will be illustrated in the next subsections.

II. Additionally to the resonance/meson decay channels the
photons can be produced directly in elementary collisions of particles.
For the present study we consider the direct photon production  by
the scattering processes
\begin{eqnarray}
\pi \pi \rightarrow \rho \gamma , \label{pipi}\\
\pi \rho \rightarrow \pi \gamma ,  \label{pirho}
\end{eqnarray}
accounting for all possible charge combinations.
We note, that we discard  $\gamma$-production in rescattering
processes with strange mesons (such as $\pi +K^*\to K+\gamma, \ \pi
+K\to K^* +\gamma, \ K+K^* \to \pi + \gamma$ etc.), since at SPS energies
the strange meson density is subdominant. However, as indicated in Ref.
\cite{Rapp04gam} such processes may play a role at higher energies.
In the next subsections we will define the contributions of the processes
$\pi \pi \rightarrow \rho \gamma , \pi \rho \rightarrow \pi \gamma $
explicitly.

III. The photon production from  meson-meson bremsstrahlung might play
an important role, too, as has been pointed out firstly by Haglin
\cite{Haglin} and later on by Liu and Rapp \cite{LiuRapp07}. These
authors have calculated the thermal photon emission rate from the
proceses $m+m \to m+m+\gamma$ and found that the bremsstrahlung
radiation is one of the dominant channels for low energy photon
production.

In the next subsections we define explicitly the  treatment of
the dominant hadronic process for direct photon production.

\subsection{Direct photon production in $\pi \pi \rightarrow \rho \gamma$
reactions}

Since at SPS energies the pion density is relatively high,  the
$\pi\pi$ annihilation channel has to be  accounted for in the
direct photon production. With increasing energies (and pion densities)
the importance of this channel grows accordingly.

The basic form of the cross section for the $\pi \pi \rightarrow \rho
\gamma$ reaction has been adopted from  Kapusta et al. in Ref.
\cite{Kapusta} where the high-energy photon production has been
evaluated in the hot hadron gas model.  However, in Ref. \cite{Kapusta}
the final $\rho$ meson has been considered on-shell, i.e. at the pole
mass $M_0=0.77$ GeV.
Since in HSD the vector mesons are treated with the full off-shell
spectral function, which depends on density and momentum
\cite{Brat07off}, some modification of the formulae from Ref. \cite{Kapusta}
has to be done in order to account for the broad mass distribution of
the final $\rho$ meson. The simplest way is to 'fold' the
cross section from Ref. \cite{Kapusta} with the mass and density
dependent spectral function of the $\rho$-mesons.

We note, that for the present study we use the same form of spectral
functions for the vector mesons as modeled in Ref. \cite{Brat07off}. In
particular, the spectral function of a vector meson $V$ with mass
$M$ at nucleon density $\rho_N$ is taken in the Breit-Wigner form:
\begin{eqnarray}
A_V(M,\rho_N) = C_1\cdot \frac{2}{\pi} \ \frac{M^2 \ \Gamma_V^*(M,\rho_N)}
{(M^2-M_{0}^{*^2}(\rho_N))^2 + (M {\Gamma_V^*(M,\rho_N)})^2},
\label{spfunV}
\end{eqnarray}
with the normalization condition for any $\rho_N$:
\begin{eqnarray}
\int\limits_{M_{min}}^{M_{lim}} A_V(M,\rho_N) \ dM =1,
\label{SFnorma}\end{eqnarray}
where $M_{lim}=2$~GeV is chosen as an upper limit for the numerical
integration. The lower limit of the vacuum $\rho$ spectral function
corresponds to the $2\pi$ decay $M_{min}=2 m_\pi$, whereas for the
in-medium collisional broadening case $M_{min}=2 m_e \to 0$ with $m_e$
denoting the electron mass. $M_0^*$ is the pole mass of the vector
meson spectral function which is $M_0^*(\rho_N=0)=M_0$ in vacuum,
however, shifted in the medium for the dropping mass scenario according
to the Hatsuda and Lee \cite{H&L92} or Brown/Rho scaling
\cite{BrownRho}. Futhermore, the vector meson width has been implemented as:
\begin{eqnarray}
\Gamma^*_V(M,|\vec p|,\rho_N)=\Gamma_V(M) + \Gamma_{coll}(M,|\vec
p|,\rho_N) . \label{gammas}
\end{eqnarray}
Here $\Gamma_V(M)$ is the total width of the vector mesons ($V=\rho,\omega$)
in the vacuum. For the $\rho$ meson we use
\begin{eqnarray}
\Gamma_\rho(M) &\simeq& \Gamma_{\rho\to\pi\pi}(M) =  \Gamma_0
\left(\frac{M_0}{M}\right)^2 \left(\frac{q}{q_0}\right)^3 \ F(M)
\label{Widthrho} \\
&& q = \frac{(M^2-4m_\pi^2)^{1/2}}{2}, \
  \ q_0 = \frac{ (M_0^2-4m_\pi^2)^{1/2} }{2}. \nonumber
\end{eqnarray}
In (\ref{Widthrho}) $\Gamma_0$ is the vacuum width
of the $\rho$-meson at the pole mass $M_0$;
$F(M)$ is a form factor taken from Ref. \cite{Rapp} as
\begin{eqnarray}
F(M)={\left( \frac{2\Lambda^2 +M_0^2}{2\Lambda^2 + M^2} \right)^2}
\label{Frapp}\end{eqnarray}
with a cut-off parameter $\Lambda=3.1$~GeV. This form factor was
introduced in Ref.  \cite{Rapp} in order to describe the $e^+e^-$
experimental data.  For the $\omega$ meson a constant total vacuum
width is used:  $\Gamma_\omega\equiv \Gamma_\omega(M_0)$, since the
$\omega$ is a narrow resonance in vacuum.

The collisional width in (\ref{gammas}) is approximated as
\begin{eqnarray}
\Gamma_{coll}(M,|\vec p|,\rho_N) = \gamma \ \rho_N < v \
\sigma_{VN}^{tot} > \approx  \ \alpha_{coll} \
\frac{\rho_N}{\rho_0} . \label{dgamma}
\end{eqnarray}
Here $v=|{\vec p}|/E; \ {\vec p}, \ E$ are the
velocity, 3-momentum and energy of the vector meson in the rest frame
of the nucleon current and $\gamma^2=1/(1-v^2)$.
Furthermore, $\sigma_{VN}^{tot}$ is the meson-nucleon total cross section.

As discussed in Ref. \cite{Brat07off} -- in order to simplify the actual
calculations -- the coefficient $\alpha_{coll}$ has been extracted in the
HSD transport calculations from the vector-meson collision rate in
$A+A$ reactions as a function of the density $\rho_N$. The numerical
results for $\Gamma_{coll}(\rho_N)$ then have been divided by
$\rho_N/\rho_0$ to fix the coefficient $\alpha_{coll}$ in
(\ref{dgamma}).  We obtain  $\alpha_{coll} \approx 150$~MeV for the
$\rho$ and $\alpha_{coll} \approx 70$~MeV for $\omega$ mesons which are
consistent with the experimental analysis in Ref. \cite{Metag07}.  In
this way the average effects of collisional broadening are incorporated
and allow for an explicit representation of the vector-meson spectral
functions versus the nuclear density, $\rho_N$.

We mention that also finite temperature effects lead to a sizable
broadening of the vector mesons spectral functions (also at baryon
chemical potential $\mu_B=0$). This is essentially due to scattering
with mesons which may contribute to the total width by 70-80 MeV at
a temperature of $\sim 170$~MeV according to the early work by Haglin
\cite{Haglin95}.  But for the present study we use a simplified
modeling of the collisional broadening width which discard an explicit
consideration of such temperature effects.  However,
the temperature effects are partly accounted here due
to explicit meson-meson interactions which also lead to changes in
the vector meson mass distribution. Since we find that the
spectral broadening of the mesons is practically not visible in the
final photon spectra,  an explicit consideration of temperature effects
is beyond the scope of this study.

In order to explore the observable consequences of vector meson
mass shifts at finite nuclear density -- as indicated by the
CBELSA-TAPS data \cite{tapselsa} for the $\omega$ meson -- the
in-medium vector meson pole masses are modeled (optionally)
according to the Hatsuda and Lee \cite{H&L92} or Brown/Rho scaling
\cite{BrownRho} as
\begin{eqnarray}
\label{Brown}
M_0^*(\rho_N)= \frac{M_0} {\left(1 + \alpha {\rho_N / \rho_0}\right)},
\end{eqnarray}
where $\rho_N$ is the nuclear density at the resonance decay position
$\vec r$, $\rho_0 = 0.16 \ {\rm fm}^{-3}$ is the normal nuclear density
and $\alpha \simeq 0.16$ for the $\rho$ and $\alpha \simeq 0.12$ for
the $\omega$ meson \cite{Metag07}. The parametrization (\ref{Brown})
may be employed also at much higher collision energies (e.g. FAIR and
SPS) and one does not have to introduce a cut-off density in order to
avoid negative pole masses. Note that (\ref{Brown}) is uniquely fixed
by the 'customary' expression $M_0^*(\rho_N) \approx M_0 (1 - \alpha
\rho_N/\rho_0)$ in the low density regime.

\begin{figure}[!]
\centerline{
\includegraphics[width=80mm]{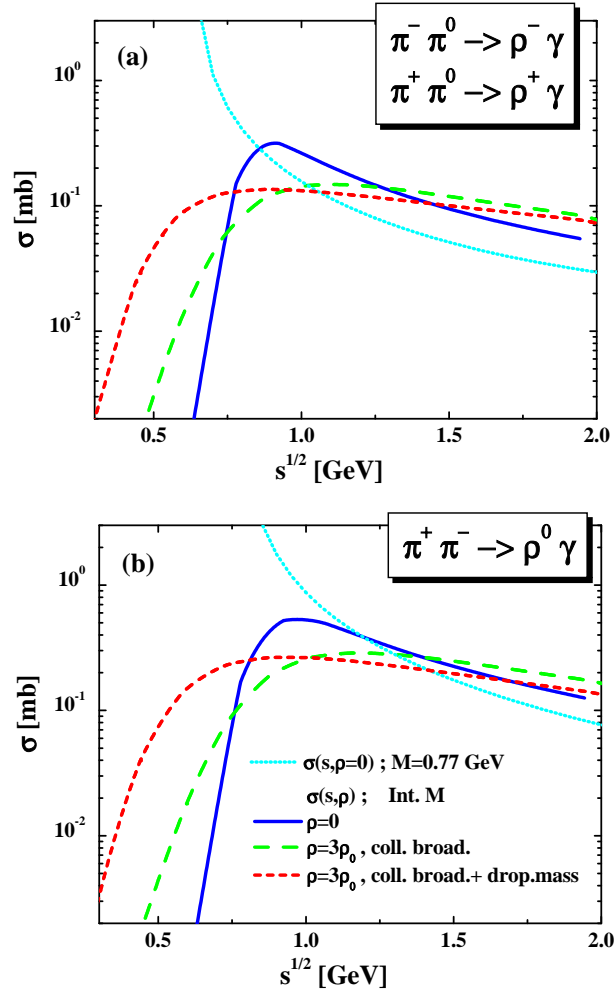}}
\caption{(Color online) The $\gamma$-production cross section
$\sigma(s,\rho)$ for the
$\pi^- + \pi^0 \to \rho^- + \gamma$,
$\pi^+ + \pi^0 \to \rho^+ + \gamma$  (upper part - (a)) and
$\pi^+ + \pi^- \to \rho^0 + \gamma$  (lower part - (b))
reaction for $\rho_N=0$ (solid lines) and in the case of
the 'collisional broadening' scenario (dashed lines) and the 'dropping
mass + collisional broadening' scenario (short dashed lines) for nuclear
density $\rho_N=3\rho_0$.
The dotted lines correspond to the results from Kapusta et al.
\cite{Kapusta} for the fixed $\rho$-meson mass $M=0.77$ GeV, while
all other lines show the yield integrated over the
$\rho$-meson mass $M$.}
\label{fig:PIPI}
\end{figure}

Thus, we model the photon-$\rho$-meson production cross section in
$\pi\pi$  reactions in the following way:  The total cross section
$\sigma_{\pi\pi \rightarrow \rho\gamma }(s,\rho_N)$ is
\begin{eqnarray}
\sigma_{\pi\pi \rightarrow \rho\gamma }(s,\rho_N)
=\int\limits_{M_{min}}^{M_{max}} dM \
\frac{d \sigma_{\pi\pi \rightarrow \rho\gamma }(s,M,\rho_N)}{d M} .
\label{xs_pipiVVtot}
\end{eqnarray}

The mass differential cross section is approximated by
\begin{eqnarray}
\frac{d \sigma_{\pi\pi \rightarrow \rho\gamma }(s,M,\rho_N)}{d M}
\: = \: \sigma_{\pi\pi \rightarrow \rho\gamma }^{0}(s,M) \cdot
A(M,\rho_N)\cdot \frac{\int\limits_{M_{min}}^{M_{max}} A(M,\rho_N) dM}
{\int\limits_{M_{min}}^{M_{lim}} A(M,\rho_N) dM },
\label{xs_pipiVV}
\end{eqnarray}
where $A(M,\rho_N)$ denotes the meson spectral function (\ref{spfunV})
for given total width $\Gamma_V^*$ (\ref{gammas}); $M_{max}=\sqrt{s}$
is the maximal kinematically allowed invariant mass of the $\rho$
meson. In Eq. (\ref{xs_pipiVV}) $\sigma_{\pi\pi \rightarrow \rho\gamma
}^{0}(s,M)$ is the vacuum cross section from Kapusta et al.
\cite{Kapusta} where the $\rho$-meson mass is considered as a free
variable (i.e. not fixed to 0.77 GeV as in \cite{Kapusta}).  Thus,
formula (\ref{xs_pipiVV}) can be used to model the vector meson
production in $\pi\pi$ reactions in the vacuum and in the medium, too.

Fig.~\ref{fig:PIPI} shows the $\gamma$-production cross section
$\sigma(s,\rho)$ for the
$\pi^- + \pi^0 \to \rho^- + \gamma$,
$\pi^+ + \pi^0 \to \rho^+ + \gamma$  (upper part) and
$\pi^+ + \pi^- \to \rho^0 + \gamma$  (lower part)
reaction for $\rho_N=0$ (solid lines) and in case of
the 'collisional broadening' scenario (dashed lines) and the 'dropping
mass + collisional broadening' scenario (short dashed lines) for nuclear
density $\rho_N=3\rho_0$.
The dotted lines correspond to the original cross section
from Kapusta et al. \cite{Kapusta} for the fixed $\rho$-meson mass
$M=0.77$ GeV, while all other lines show the cross section integrated over
the mass $M$.
Note that the cross section from Kapusta et al. \cite{Kapusta}
for the fixed $\rho$-meson mass  diverges at threshold.
However, the folding over the spectral function of the $\rho$
meson leads to an endothermic behavior of the cross section
(cf. the solid line) instead of an exothermic (dotted line).

As seen from Fig. \ref{fig:PIPI} the cross section in the
'collisional broadening' scenario is basically smeared out close to
threshold.  Only when incorporating additionally a dropping mass the
thresholds are shifted down in energy such that the production cross
sections become enhanced in the subthreshold regime with increasing
nuclear density $\rho_N$.

%
\subsection{Direct photon production in $\pi \rho \rightarrow \pi \gamma$
reactions}

\begin{figure}[!]
\centerline{
\includegraphics[width=80mm]{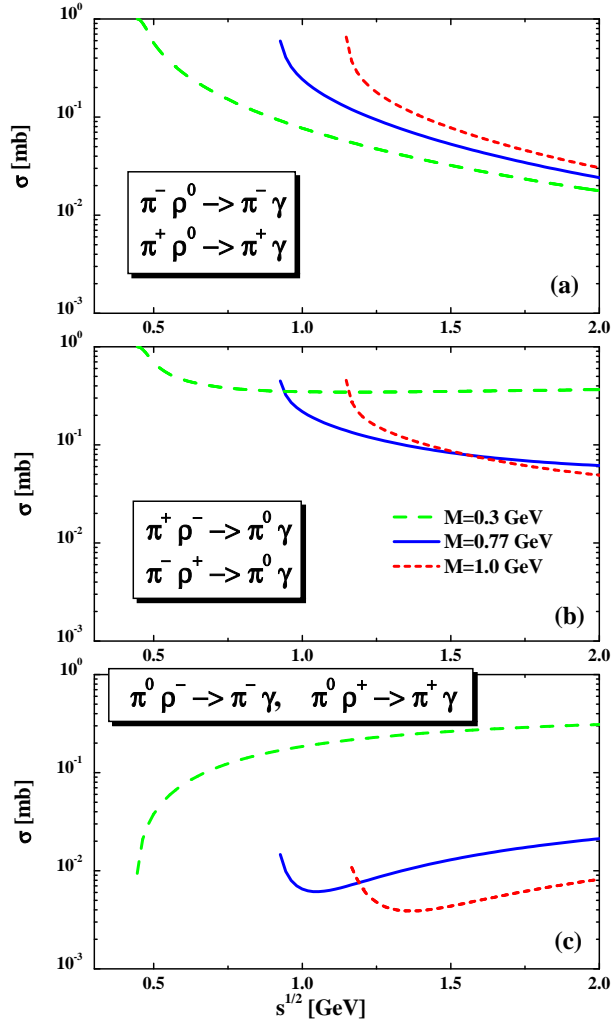}}
\caption{(Color online) The production cross section $\sigma(s)$ for the
$\pi^- + \rho^0 \to \pi^- + \gamma$,
$\pi^+ + \rho^0 \to \pi^+ + \gamma$  (upper part - (a)), \
$\pi^+ + \rho^- \to \pi^0 + \gamma$,
$\pi^- + \rho^+ \to \pi^0 + \gamma$  (middle part - (b)) and
$\pi^0 + \rho^- \to \pi^- + \gamma$,
$\pi^0 + \rho^+ \to \pi^+ + \gamma$  (lower part - (c))
reaction.
The dashed lines corresponds to the $\rho$-meson mass
$M=0.3$ GeV, the solid lines stand for $M=0.77$ GeV
while the short dashed lines correspond to $M=1.0$ GeV.}
\label{fig:pirho}
\end{figure}

In Fig. \ref{fig:pirho} we show the production cross section
$\sigma(s)$ for the
$\pi^- + \rho^0 \to \pi^- + \gamma$,
$\pi^+ + \rho^0 \to \pi^+ + \gamma$  (upper part), \
$\pi^+ + \rho^- \to \pi^0 + \gamma$,
$\pi^- + \rho^+ \to \pi^0 + \gamma$  (middle part) and
$\pi^0 + \rho^- \to \pi^- + \gamma$,
$\pi^0 + \rho^+ \to \pi^+ + \gamma$  (lower part)
reactions.
Here again we use the vacuum cross sections from Kapusta et al.
\cite{Kapusta} and consider the $\rho$-meson mass as
a free variable since the production of the vector mesons in HSD
is realized with respect to the spectral function $A(M,\rho_N)$.
The dashed lines corresponds to the $\rho$-meson mass
$M=0.3$ GeV, the solid lines stand for $M=0.77$ GeV,
while the short dashed lines correspond to $M=1.0$ GeV.

As follows from Fig. \ref{fig:pirho} the $\pi+\rho\to \pi+\gamma$
cross section depends very strongly on the mass of the initial $\rho$ meson
-- for low $M$ the threshold is shifted to low $\sqrt{s}$.
Thus, one can expect an enhancement of  $\gamma$ production in case of
this channel for the in-medium scenarios due to the
enhanced population of the low mass $\rho$ mesons.

\subsection{Photon production by the decay
$a_1 \rightarrow \pi \gamma$}

The photons can be also emitted from the decay $a_1 \to \pi \gamma$.
In spite that the branching ratio of this process is
not well known experimentally  and expected to be very small
(we use $Br(a_1 \to \pi \gamma) \sim 1.5\cdot 10^{-3}$ \cite{Zilens84}),
this contribution is also considered in our investigation.
The study of the $a_1$ dynamics is an interesting problem itself since
it is related to chiral symmetry restoration at high densities and
temperatures as pointed out in
Refs. \cite{Koch_a1,Rapp_a1,Leupold_a1,Vogel07_a1}.

The production of $a_1$ mesons in HSD stems from $BB$ and $mB$
collisions via string excitation and decay at high energies and
by $\pi+\rho \leftrightarrow a_1$ reactions at low energies. The mass of
the $a_1$ meson is distributed according to the Breit-Wigner
spectral function - Eq. (\ref{spfunV}) - with a mass ($\mu$) and
density ($\rho_N$) dependent width $\Gamma_{a_1}^*$.
The total $a_1$ width $\Gamma_{a_1}$ can be written as
\begin{eqnarray}
\Gamma^*_{a_1}(\mu,|\vec p|,\rho_N)=\Gamma_{a_1}(\mu,\rho_N)
+ \Gamma_{a_1:coll} (\mu,|\vec p|,\rho_N) ,
\label{gamma_a1}
\end{eqnarray}
where the collisional width is approximated  by
$\Gamma_{a_1:coll}(\mu,|\vec p|,\rho_N)\simeq \alpha_{a_1:coll}
\ \rho_N/\rho_0$ with $\alpha_{a_1:coll}\simeq 150$~MeV.

\begin{figure}[!]
\centerline{
\includegraphics[width=80mm]{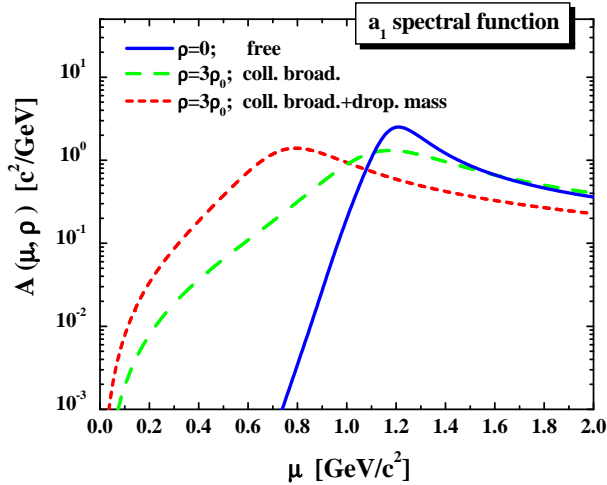}}
\caption{(Color online) The spectral function for the $a_1$ meson
for $\rho_N=0$ (solid line) and in case of the 'collisional broadening'
scenario (dashed line) and the 'dropping mass + collisional broadening'
scenario (short dashed line) for nuclear density $\rho_N=3\rho_0$.}
\label{fig:A_a1}
\end{figure}

Since the $a_1$ meson decays dominantly to $\pi+\rho$ \cite{Leupold_a1}
one has to take into account the mass distribution of the final $\rho$
meson $A(M,\rho_N)$ when calculating the width of the $a_1$ meson.
Thus formula (\ref{Widthrho}) for the $a_1$ width
$\Gamma_{a_1}(\mu,\rho_N)$ has to be modified accordingly:
\begin{eqnarray}
&&\Gamma_{a_1}(\mu,\rho_N) = \Gamma_{a_1}^0 \
\left(\frac{\mu_0}{\mu}\right)^2    \
\frac{ \int\limits_{M_{min}}^{\mu -m_\pi} dM \ q^3(\mu,M)
\ A(M,\rho_N) }
{ \int\limits_{M_{min}}^{\mu_0-m_\pi} dM \ q_0^3(\mu_0,M)
 \ A(M,\rho_N)} ;
					\label{Widtha1} \\
&& q(\mu,M) = \frac{\sqrt{\lambda(\mu,M,m_\pi)}}{2\mu}, \ \ \
 q_0(\mu_0,M) = \frac{\sqrt{\lambda(\mu_0,M,m_\pi)}}{2\mu_0} .
\nonumber
\end{eqnarray}
Here $\lambda(x,y,z) = (x^2-(y-z)^2)(x^2-(y+z)^2)$
while $m_\pi$ is the pion mass;
$\mu_0, \Gamma_{a_1}^0$ are the vacuum pole mass
and width of the $a_1$ spectral function with $\Gamma_{a_1}^0$ taken
as 400 MeV.

According to Eq. (\ref{Widtha1}) the $a_1$ width depends on the
 $\rho$ meson spectral function $A(M,\rho_N)$ and,
thus, is sensitive to the $\rho$ meson in-medium effects.
We stress, that such a 'coupling' of the $a_1$ and $\rho$ spectral
functions has to be accounted for when drawing conclusions about an
experimental observation of chiral symmetry restoration by
measuring the $a_1$ meson properties in heavy-ion experiments (e.g. the
corresponding results in Ref. \cite{Vogel07_a1} should be strongly
effected).

Figure \ref{fig:A_a1} shows the spectral function
$A_{a_1}(\mu,\rho_N)$ for the $a_1$ meson for $\rho_N=0$ (solid line)
and in the case of the 'collisional broadening' scenario (dashed line)
and the 'dropping mass + collisional broadening' scenario (short dashed
line) for nuclear density $\rho_N=3\rho_0$.  Here the in-medium
scenarios are related to both the $a_1$ as well as the $\rho$ meson.

\begin{figure}[t]
\centerline{
\includegraphics[width=80mm]{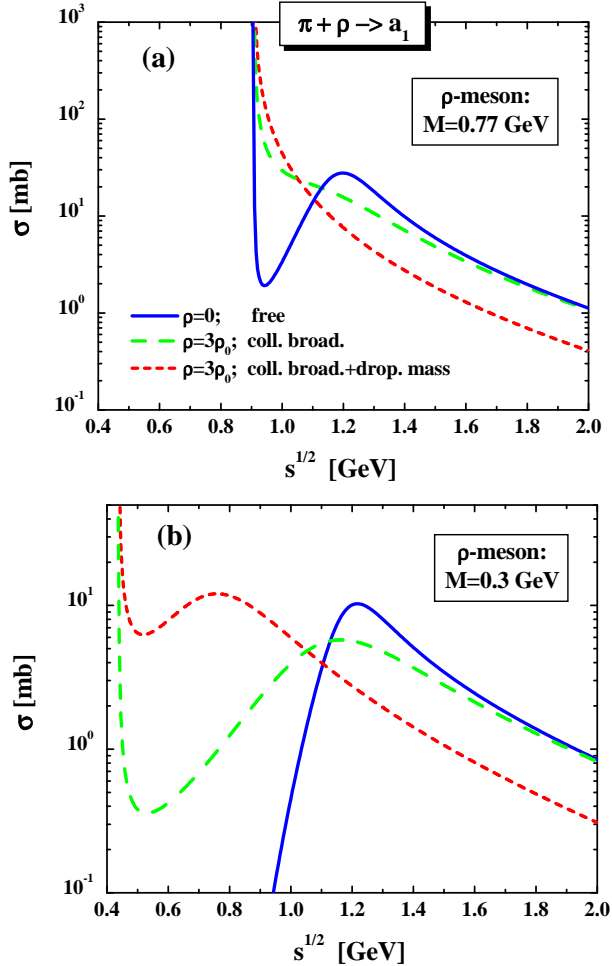}}
\caption{(Color online) The production cross section $\sigma(s,M)$
for the $\pi + \rho \to a_1$ reaction for $\rho_N=0$ (solid lines)
and in the case of the 'collisional broadening' scenario
(dashed lines) and the 'dropping mass + collisional broadening' scenario
(short dashed lines) for nuclear density $\rho_N=3\rho_0$.
The upper part (a) corresponds to $M=0.77$
GeV while the lower part (b) shows the results for $M=0.3$ GeV. }
\label{fig:xs_a1}
\end{figure}

The production cross section $\sigma(s,M)$ for the
$\pi + \rho \to a_1$ reaction (for all possible charge combinations)
is calculated as follows,
\begin{eqnarray}
\sigma_{\pi+\rho \to a_1}(s,M,\rho_N) = \frac{6\pi^2}{q^2(s,M)} \
\Gamma_{a_1}(s,\rho_N) A_{a_1}(s,\rho_N),
\label{eq:xs_a1}
\end{eqnarray}
where $s\equiv \mu^2$ and the width $\Gamma_{a_1}(s,\rho_N)$
and momentum $q(s,M)$ are defined by Eq. (\ref{Widtha1}).

In Fig. \ref{fig:xs_a1} we display the production cross section
$\sigma(s,M)$ for the $\pi + \rho \to a_1$ reaction for $\rho_N=0$ (solid
lines) and in case of the 'collisional broadening' scenario (dashed
lines) and the 'dropping mass + collisional broadening' scenario (short
dashed lines) for nuclear density $\rho_N=3\rho_0$. The upper part
corresponds to the $\rho$ meson mass of $M=0.77$ GeV while the lower
part shows the results for $M=0.3$ GeV.  As seen from Fig.
\ref{fig:xs_a1}  the $a_1$ production cross section in the $\pi+\rho$
reaction is very sensitive to the in-medium scenarios  as
well as to the mass $M$ of the initial $\rho$ meson (even in the free
case -- cf. solid lines).

\subsection{Direct photon production from meson-meson bremsstrahlung}

The implementation of photon bremsstrahlung from hadronic reactions in
transport approaches is based on the 'soft photon' approximation.
The soft-photon approximation (SPA) \cite{GaleK87} relies on the
assumption that the radiation from internal lines is negligible and the
strong interaction vertex is on-shell. In this case the strong
interaction part and the electromagnetic part can be separated,
so the soft-photon cross section for the reaction
$1+2\to  1 + 2 + \gamma$ can be written as
\begin{eqnarray}
&& q_0\frac{d^3 \sigma^\gamma}{d^3 q} = \frac{\alpha}{4 \pi}
\frac{{\bar \sigma (s)}}{q_0^2}
			\label{brems} \\
&& {\bar \sigma(s)} = \frac{s - (M_1 + M_2)^2}{2 M_1^2} \sigma(s), \nonumber
\end{eqnarray}
where $M_1$ is the mass of the charged accelerated particle,
$M_2$ is the mass of the second particle; $q_0, q$ are the energy
and momentum of the photon.
In (\ref{brems})  $\sigma(s)$ is the on-shell elastic cross section
for the reaction $1+2\to 1+2$. This approximation has also been
employed by Haglin in Ref. \cite{Haglin}.

In Ref. \cite{LiuRapp07} the photon bremsstrahlung from $\pi+\pi$ and
$\pi+K$ elastic collisions has been calculated using the $U_{em}(1)$-gauged
meson-exchange model which includes the photon coupling to pseudoscalar
and vector mesons. Indeed, such calculations go beyond the SPA model.
However, the direct comparison of SPA and the $U_{em}(1)$ models -
cf. Fig. 4 in Ref. \cite{LiuRapp07} - show a very good agreement
between the models, which allows us to use the simplified
SPA formula for our purpose here.
Thus, we have calculated the photon bremsstrahlung from all elastic
meson-meson scattering $m_1 + m_2 \to m_1 + m_2 + \gamma$ (where $m =
\pi, \eta, K, \bar K, K^0, K^*, \bar K^*, K^{*0}$), which occur during
the heavy-ion collisions by applying the SPA formula (\ref{brems}).

\section{Results for the photon production in A+A collisions at
SPS energies}

Now we step on to the description of direct photon production
from heavy-ion collisions applying the HSD transport model incorporating
all photon production channels from elementary reactions as described in
the previous Section.

\begin{figure}[!]
\centerline{\psfig{figure=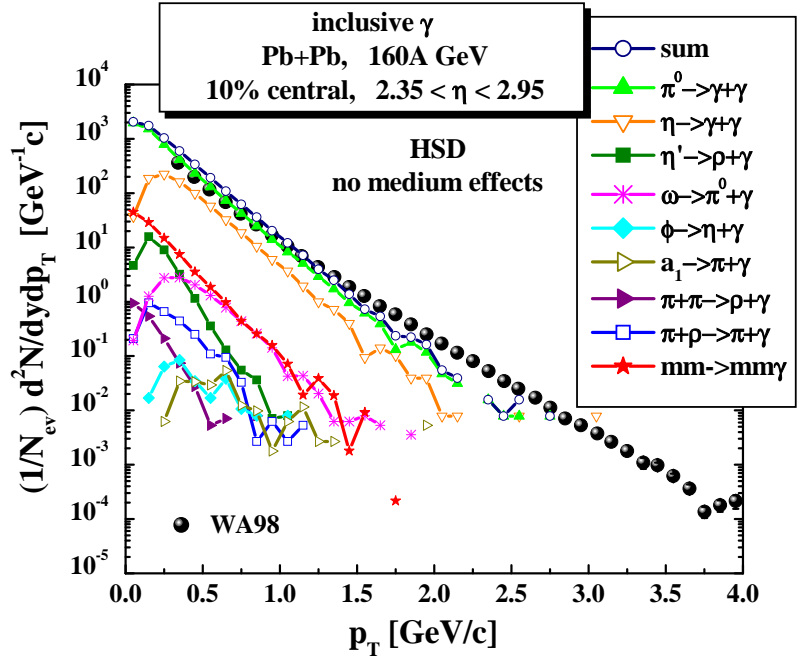,width=10.5cm}}
\caption{(Color online) The inclusive photon transverse momentum distribution
   for  central 158~{\it A}~GeV $^{208}$Pb\/+\/$^{208}$Pb
   collisions at $2.35 \le \eta \le 2.95$.
The solid symbols and arrows corresponds to the WA98
data \protect\cite{WA98data},
the lines correspond to the HSD results for the various channels
as indicated in the legend (without medium effects for the vector
mesons).}
\label{fig:WA98incl}
\end{figure}

In Fig. \ref{fig:WA98incl} we start with a comparison of the HSD
results for the inclusive photon transverse momentum distribution for
10\% central 158~{\it A}~GeV $^{208}$Pb\/+\/$^{208}$Pb collisions with
the experimental data from the WA98 Collaboration \cite{WA98data}
(solid dots). We select photons in the pseudorapidity interval $2.35
\le \eta \le 2.95$ in the laboratory frame which corresponds to
mid-rapidity in the center-of-mass frame. Our calculations show that
such cut in pseudorapidity reduces the photon yield by about of factor
5, however, practically doesn't change the shape of the $p_T$ spectra.
The WA98 data correspond to the transverse momentum interval $0.5\le
p_T \le 4$ GeV$/c$ while the HSD calculation includes also very low
$p_T$ momenta but extends up to 2.5-3 GeV$/c$ only since the high $p_T$
tail is very hardly reachable due to limitied statistics.

Fig. \ref{fig:WA98incl} presents the HSD calculations without medium
effects for the vector mesons (the definition of the lines are
indicated in the legend).  As seen from Fig. \ref{fig:WA98incl} the
dominant channels for the inclusive photon production are the photon
from $\pi$ and $\eta$ decays ($\pi\to \gamma+\gamma$ and $\eta \to
\gamma +\gamma$), whereas other channels are down by more than order of
magnitude.  The HSD results agree very well with the experimental data
which is, indeed, expectable since the HSD model provides a good
description of the pion transverse momentum spectra at SPS energies
\cite{Brat03} and predicts a meson $m_T$ scaling \cite{Cass01mt}.

In order to obtain the information about the direct photon production,
one needs to subtract the 'background' contributions which are
dominated by the mesonic decay processes. While the $\pi^0$ and $\eta$
spectra are measured directly by the same experiments \cite{WA98data},
it is possible to estimate their photon decay in a reliable way.
Indeed, the life times of $\pi$'s and $\eta$'s are large such that they
basically decay at the end of reaction stage, i.e. in vacuum.
However, the situation with the $\eta^\prime \to \rho+\gamma,
\omega\to \pi^0+\gamma, \phi\to \eta+\gamma, a_1\to \pi+\gamma$ decays
are not so transparent since these mesons cannot be mesured directly
by the WA98 detector.

\begin{figure}[!]
\centerline{\psfig{figure=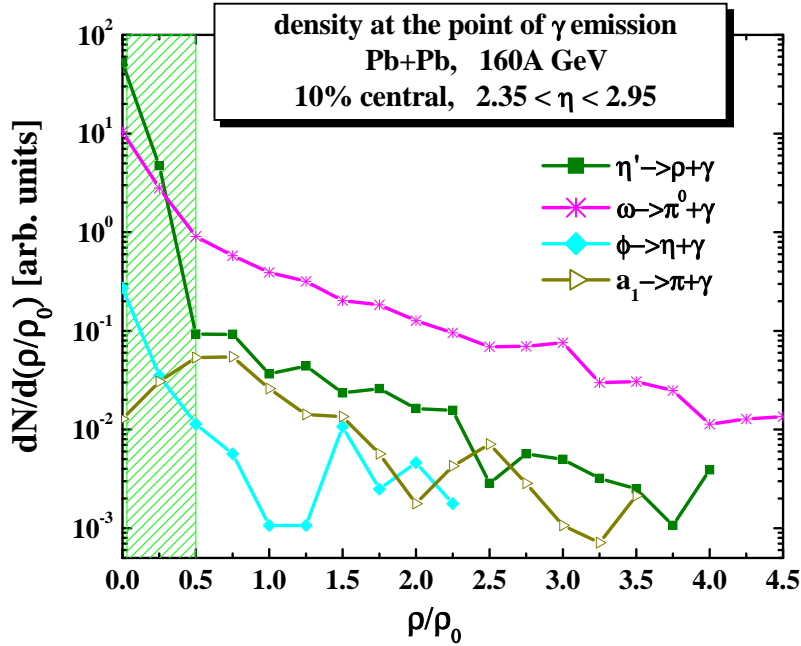,width=10.5cm}}
\caption{(Color online) The nucleon density distribution at the photon emission point
(in units of $\rho_0$)
   for  central 158~{\it A}~GeV $^{208}$Pb\/+\/$^{208}$Pb
   collisions at $2.35 \le \eta \le 2.95$.
The definition of the lines is indicated in the legend.}
\label{fig:dens}
\end{figure}

The WA98 Collaboration has subtracted the contribution of
hadronic decays by calculations based on the assumption of
$m_T$-scaling with the same slope of the $m_T$-spectra as measured in the
$\pi^0$ spectrum and with relative normalizations $R_{hadron/\pi^0}$
(equivalent to the asymptotic ratios for $p_T \to \infty$) \cite{WA98data}.
However, such a procedure is a rought approximation especially for the
short living resonances which can emit photons inside the hot
fireball shortly after creation and, thus, can not be reconstruct
in experiment. This statement is illustrated in Fig.
\ref{fig:dens} which shows the density distribution at the photon
emission points from $\eta^\prime, \omega, \phi, a_1$ decays for central
158~{\it A}~GeV $^{208}$Pb\/+\/$^{208}$Pb collisions at $2.35 \le \eta
\le 2.95$. One can see from Fig. \ref{fig:dens} that there is some
fraction of $\omega, \phi, a_1$ mesons that emit photons for $\rho_N
\ge \rho_0 /2$ (e.g.  $\sim 25$\% of $\omega$'s and $\sim 70$\% of
$a_1$'s), whereas a relatively long leaving $\eta^\prime$ meson
decays dominantly at low densities.

\begin{figure}[!t]
\phantom{a}\hspace*{-4.5cm} \psfig{figure=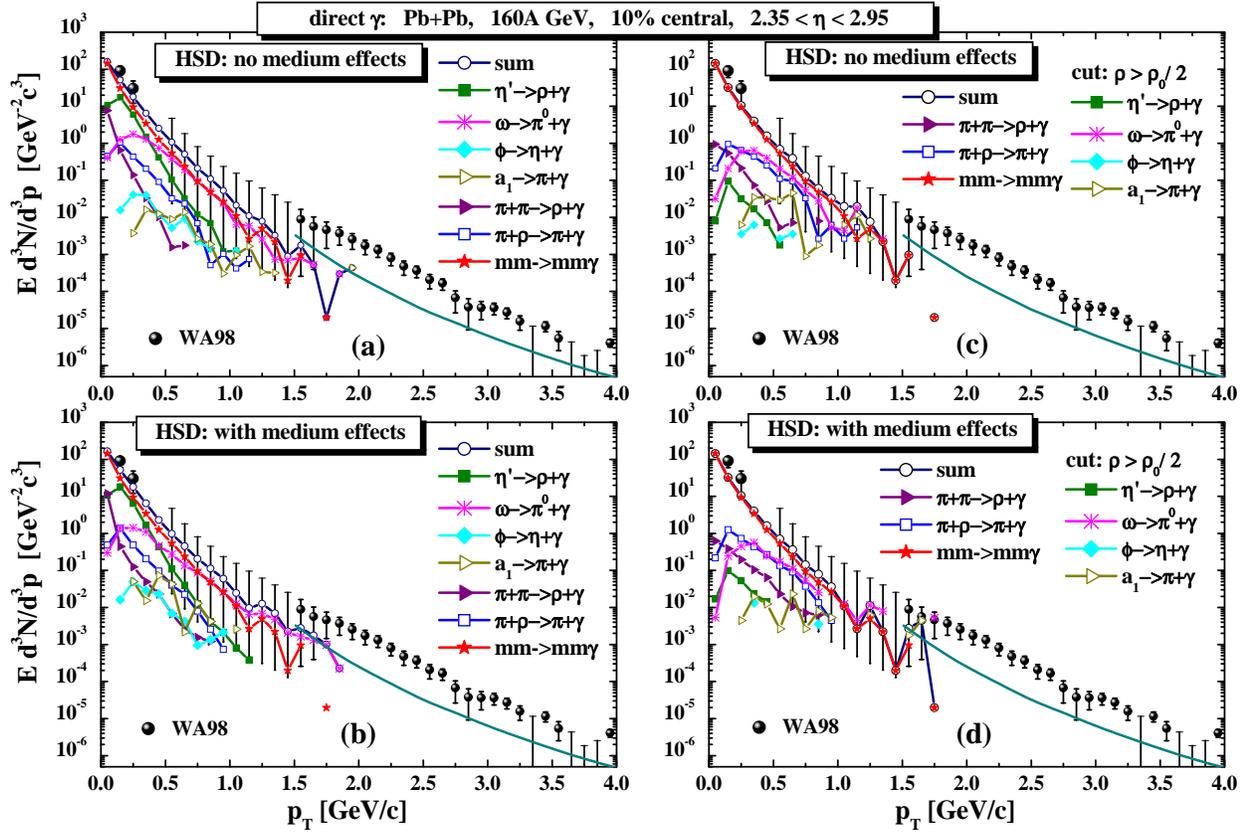,width=12cm}
\caption{(Color online) The invariant direct photon multiplicity
   for  central 158~{\it A}~GeV $^{208}$Pb\/+\/$^{208}$Pb
   collisions at $2.35 \le \eta \le 2.95$.
The solid symbols and arrows corresponds to the WA98
data \protect\cite{WA98data},
the lines stand for the HSD calculations without (upper row - (a),(c)) 
and with (lower row - (b),(d)) medium effects (i.e. collisional broadening)
for the vector mesons.
The definition of the lines is indicated in the legend.
The solid line for $p_T>1.5$ GeV$/c$ gives the contribution from
the prompt photons.
The right panel (c),(d) shows the spectra including the contribution of the
$\eta^\prime, \omega, \phi, a_1$ photon decay only for $\rho_N > \rho_0/2$
whereas the left panel (a),(b) shows the total contribution of all hadronic 
channels at all densities.}
\label{fig:WA98data}
\end{figure}

In order to demonstrate the possible contribution
of the in-medium decays of $\eta^\prime, \omega, \phi, a_1$ mesons
to the direct photon spectra reported by the WA98 Collaboration
we show in the r.h.s of Fig. \ref{fig:WA98data} the invariant mid-rapidity
spectra including the contribution of the
$\eta^\prime, \omega, \phi, a_1$ photon decay only for $\rho_N > \rho_0/2$
whereas the l.h.s. of Fig. \ref{fig:WA98data} shows the
total contribution of all hadronic channels at all densities (for
comparison). Indeed, such a cut in density leads to a
substantial reduction of the hadronic decay 'background'
especially for low $p_T$, however, it becomes essential
at $p_T \le 0.7$ GeV/c. Thus, our calculations provide the
scale of possible uncertainties in the experimental background subtraction
which is important for the physical interpretation of the
experimental results.

One can see from Fig. \ref{fig:WA98data} that the dominant channel for
low $p_T$ photon production is  meson-meson bremsstrahlung. This
is in line with the analysis by Haglin \cite{Haglin} and
Liu and Rapp \cite{LiuRapp07}.
The contributions of the other channels such as direct photon production by
meson-meson collisions $\pi +\pi\to \rho +\gamma$ and
$\pi +\rho\to \pi +\gamma$ are found to be suppressed relative to the
meson-meson bremsstrahlung. Also we have
investigated the influence of in-medium effects such as collisional
broadening for the vector meson spectral functions which leads (e.g.)
to enhancement of low mass dilepton production
\cite{Brat07off,Brat08SPS}. However, we here obtain only a small impact
on the direct photon production which is hidden below the dominant
bremsstrahlung contribution. The same holds also for other
in-medium scenario such as a dropping mass of the vector meson.

\begin{figure}[!t]
\centerline{\psfig{figure=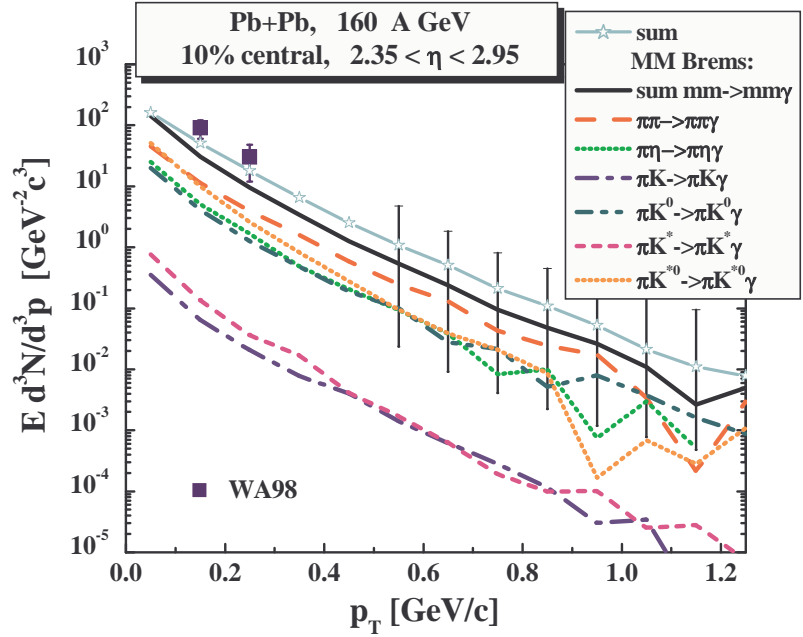,width=10.5cm}}
\caption{(Color online) The invariant direct photon multiplicity
for  central 158~{\it A}~GeV $^{208}$Pb\/+\/$^{208}$Pb
collisions at $2.35 \le \eta \le 2.95$.
The solid symbols and arrows corresponds to the WA98
data \protect\cite{WA98data},
the lines stand for the HSD calculations: the upper solid line with the
stars shows the sum over all channels, the solid line corresponds to the
sum over all meson-meson bremsstrahlung channels which are specified
by the  various lines.
The definition of the lines is indicated in the legend.}
\label{fig:WA98dec}
\end{figure}

In Fig. \ref{fig:WA98dec} we specify the meson-meson bremsstrahlung
by showing the contributions of different sub-channels
$\pi+\pi \to \pi +\pi + \gamma$, \
$\pi+\eta \to \pi +\eta + \gamma$,  \
$\pi+K \to \pi +K + \gamma$ \ with  $K=(K^+,K^-)$, \
$\pi+K^0 \to \pi +K^0 + \gamma$, \
$\pi+K^* \to \pi +K^* + \gamma$ and
$\pi+K^{*0} \to \pi +K^{*0} + \gamma$.
We stress again that only the elastic meson-meson collisions have been
accounted here. Indeed, the final result is very sensitive to the
meson-meson elastic cross section which is, however, very hard to
measure experimentally. We have used a 10 mb elastic cross section
for all meson-meson scatterings in the present study.  As follows from
Fig. \ref{fig:WA98dec} the dominant contribution stems from $\pi+\pi$
elastic scattering; the contributions from $\pi + K^{*0}$, $\pi + K^0$
and $\pi + \eta$ are also visible at low $p_T$. This is again consistent
with Refs. \cite{Haglin,LiuRapp07} where the authors found that the
photon emission rate from pion-kaon scattering is at the $\sim 20$\%
level of the one from pion-pion scattering.

We mention that the solid line at high $p_T$ in Fig.
\ref{fig:WA98data} shows the estimated contribution from the prompt
photons. For that we have convoluted the pp data fit from
Ref.~\cite{Srivastava} with the nuclear overlap function $T_{AB}$.
Indeed, it gives a lower estimate for the prompt photons since the
nuclear effects (as e.g. Cronin effect) have been ignored.  For futher
discussions on the contributions of the prompt as well as thermal
photons we address to Refs. \cite{Srivastava,Rapp04gam} and references
therein.

\section{Results for the photon production in p+A collisions at
SPS energies}

In order to draw solid conclusions on the direct photons from
heavy-ion collisions one needs some 'reference systems', i.e. to compare
to p+p and/or p+A data. Recently the WA98 Collaboration has provided
preliminary data on p+C and p+Pb collisions at $\sqrt{s}=17.4$ GeV.
In this Section we present the HSD results for these systems.

\begin{figure}[!t]
\centerline{\psfig{figure=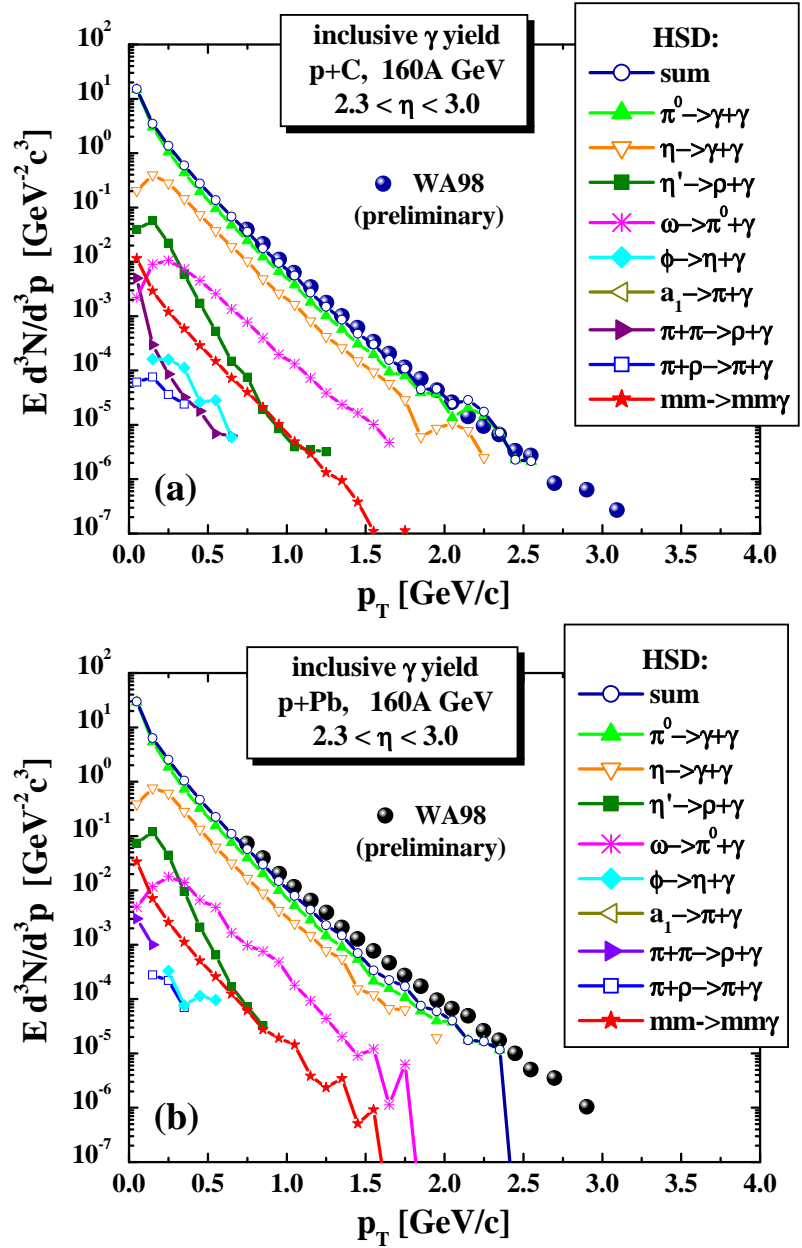,width=10.5cm}}
\caption{(Color online) The inclusive invariant photon multiplicity
   for  158~{\it A}~GeV $p + ^{12}$C (upper part - (a)) and $p +^{208}$Pb
(lower part - (b)) at $2.3 \le \eta \le 3.0$.
The solid dots corresponds to the preliminary WA98 data
\protect\cite{WA98_pA08}; the lines stand for the HSD calculations.
The definition of the lines is indicated in the legend.}
\label{fig:WA98pAincl}
\end{figure}

We start again with the inclusive invariant photon spectra
from p+C and p+Pb collisions at mid-rapidity for 160 A GeV which
are depicted in Fig. \ref{fig:WA98pAincl}.
The solid dots corresponds to the preliminary WA98 data
\cite{WA98_pA08}, the various lines stand for the HSD calculations.
Again, as in a case of Pb+Pb, the dominant contribution comes
from the $\pi^0$ and $\eta$ photonic decays. As seen from
Fig. \ref{fig:WA98pAincl} HSD provides a very good description
of the inclusive experimental data.

\begin{figure}[!]
\centerline{\psfig{figure=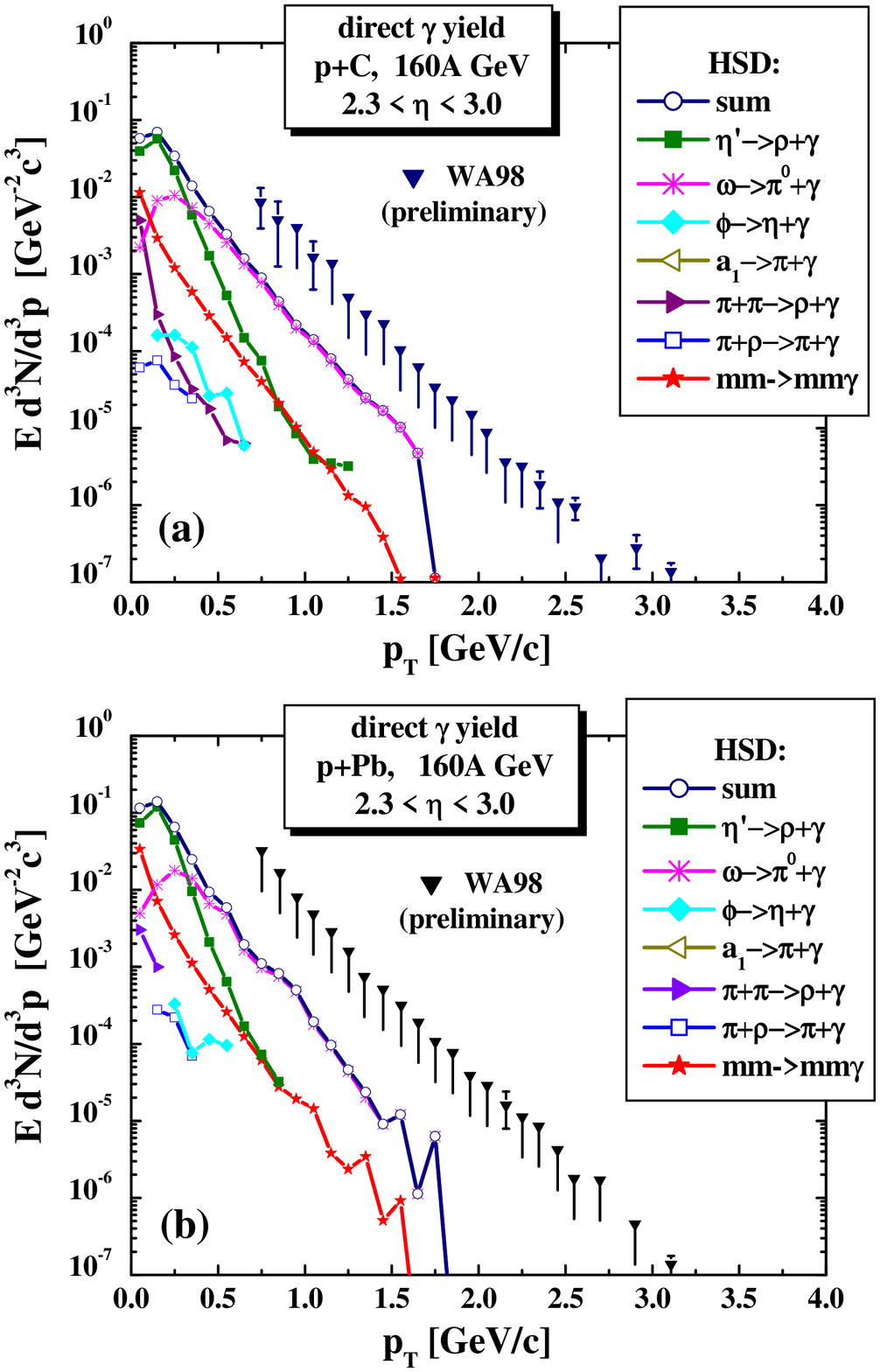,width=10.5cm}}
\caption{(Color online) The invariant direct photon multiplicity
   for  158~{\it A}~GeV $p + ^{12}$C (upper part - (a)) and $p +^{208}$Pb
(lower part - (b)) at $2.3 \le \eta \le 3.0$.
The solid dots corresponds to the preliminary WA98 data
\protect\cite{WA98_pA08}; the lines stand for the HSD calculations.
The definition of the lines is indicated in the legend.}
\label{fig:WA98pAdir}
\end{figure}

In order to focus on the direct photon contribution the WA98 collaboration
has subtracted the contributions of the hadronic decays in a similar
way as for the Pb+Pb collisions. However, due to experimental uncertainties
in the 'background ' subtraction the preliminary WA98 data
correspond to the upper limits on direct photon production
\cite{WA98_pA08} and are shown by the (down) arrows in Fig.
\ref{fig:WA98pAdir} for p+C (upper part) and p+Pb (lower part).
The various lines in Fig. \ref{fig:WA98pAdir} stand for the HSD result.
Here we do not apply any cuts in density in order to obtain an
upper estimate for the hadronic decays. Contrary to the Pb+Pb collisions
the contribution from the meson-meson bremsstrahlung is suppressed
compared to the $\eta^\prime$ and $\omega$ decay due to the small meson
density in p+A collisions.
As seen from Fig. \ref{fig:WA98pAdir} the HSD results are much below
the upper limits from WA98. Since it is unlikely that there are some new
channels for photon production in p+A reactions,
which are not visible in Pb+Pb collisions, we expect that the final data
might go down, too. Thus, it is quite important to obtain
precise p+A data from the experimental side in order to
check the theoretical models more accurately.

\section{Summary}

In this study a detailed analysis of the low $p_T$ photon production
in p+C, p+Pb and Pb+Pb collisions at 160 A GeV has been presented
within the microscopic HSD transport approach that incorporates a full
off-shell propagation of the vector mesons \cite{Brat07off}.
We stress, that the HSD approach contains only
hadronic degrees of freedom and strings and doesn't include the phase
transition from a QGP to hadronic matter explicitly; the partonic
interactions are treated in HSD only inside the strings (and in terms
of leading quarks and diquarks). However, the model has all
non-equilibrium dynamical features and demonstrates a good ability in
describing a variety observables for heavy-ion, proton- and pion-
induced reactions. Accordingly, we apply our model to study the direct
photon production from hadronic sources which dominate at low $p_T$.

In particular the following  hadronic sources for direct photon
production in heavy-ion collisions have been incorporated:

-- photon emission from  elementary  meson-meson rescatterings,
where the processes $\pi \rho \rightarrow \pi \gamma$ and $\pi \pi
\rightarrow \rho \gamma$ are  dominant due to the high pion
production rate. The novel issue here is that we accounted for the
off-shellness of the initial/final $\rho$ meson and extended the
vacuum cross sections for these processes from Kapusta et al.
\cite{Kapusta} for the in-medium case with the full off-shell $\rho$
meson spectral function. It allows us to investigate the influence
of the in-medium effects such as 'collisional broadening' on the
photon $p_T$ spectrum.

-- meson-meson bremsstrahlung from the elastic meson-meson scattering
$m_1 + m_2 \to m_1 + m_2 + \gamma$ (where $m = \pi, \eta, K, \bar K,
K^0, K^*, \bar K^*, K^{*0}$), which we accounted for by applying the
SPA formula (\ref{brems}).

We have found that the enhancement of the low $p_T$ photon emission
from elementary meson-meson rescattering  due to a collisional
broadening  of the vector-meson spectral functions is hardly visible in
the final spectra which are dominated by bremsstrahlung type processes.
Thus our non-equilibrium dynamical calculations support the early
findings in Refs. \cite{Haglin,LiuRapp07} based on the hadron gas models.

Also we have investigated the uncertainties in the extraction of the direct
photon yield from the measured inclusive photon spectra
which are dominated by the hadronic decays, in particular $\pi^0, \eta,
\eta^\prime, \omega$ and $a_1$. Here the HSD model shows a very good
agreement with the measured inclusive photon transverse
momentum distribution.
While the 'background' contributions from the $\pi^0$ and $\eta$ decay
can be subtracted in a reliable way -- since these mesons are measured by
the same WA98 experiment -- there remains a problem in the subtraction
of the contributions from short living resonances which decay inside the
hot and dense fireball.  We have estimated the possible contribution
of such in-medium decay processes by comparing two calculations --
in the first we have selected the photons coming from the mesonic
decays inside the fireball by applying a density cut $\rho_N \ge
\rho_0/2$, in the second we have accounted for the
mesonic decays over the full time history. We found that the density
cut reduces drastically the final yield which depends directly on the
actual value of the cut. However, some amount of photons from the inside
$\omega$ decay for $\rho_N \ge \rho_0/2$ is visible in the final spectra
and might be misidentified as direct photons.

For reference  we have also studied the photon production for p+C and
p+Pb collisions at 160 GeV.  Again the HSD model provides a good
description of the preliminary experimental data from the WA98 Collaboration
on the inclusive spectra, however,  is much below the upper limits
for the direct photons (even without a subtraction of $\eta^\prime,
\omega, \phi$ and $a_1$ photon decays). In this respect it will be very
helpful to have the final experimental data (with high accuracy).

\section*{Acknowledgments}
The authors acknowledge inspiring discussions with W. Cassing,
C. Gale, R. Rapp and S. Turbide.
This  work  was partially supported by INTAS, grant number 06-1000012-8914,
and the Federal agency of Russia for atomic energy (Rosatom).


\end{document}